\documentclass[12pt,a4paper]{article}
\usepackage{graphicx}  
\usepackage{amsmath,amsfonts,amssymb} 

\title{Subcritical transition to turbulence in wall-bounded flows:\\
the case of plane Poiseuille flow}

\author{Paul Manneville$^1$ and Masaki Shimizu$^2$\\
\normalsize $^1$ Hydrodynamics Laboratory, CNRS-UMR7646,\\ 
\normalsize \'Ecole Polytechnique, Palaiseau, 91128 France\\
\normalsize $^2$ Graduate School of Engineering Science,\\
\normalsize Osaka University, Toyonaka, 560-0043 Japan}
\date{\large 22\`eme Rencontre du Non Lin\'eaire, Paris, March 26--29, 2019\\
\normalsize proceedings p. 39--44}

\begin{document}

\maketitle

\begin{abstract}
In wall-bounded flows, the laminar regime remain linearly stable up to large values of the Reynolds number while competing with nonlinear turbulent solutions issued from finite amplitude perturbations.
The transition to turbulence of plane channel flow (plane Poiseuille flow) is more specifically considered {\it via\/} numerical simulations.
Previous conflicting observations are reconciled by noting that the two-dimensional directed percolation scenario expected for the decay of turbulence may be interrupted by a symmetry-breaking bifurcation favoring localized turbulent bands.
At the other end of the transitional range, a preliminary study suggests that the laminar-turbulent pattern leaves room to a featureless regime beyond a well defined threshold to be determined with precision.
   
\end{abstract}
\sloppy

\section{Context\label{mannevsect1}}
%\noindent\begin{minipage}{0.48\textwidth}
Closed systems such as Rayleigh--B\'enard convection become turbulent according to a {\it globally supercritical scenario\/} starting from an intuitively simple linear instability.
As the control parameter is increased, this {\it primary\/} instability is then followed by a small number of successive, {\it secondary\/}, {\it tertiary\/},\dots\ instabilities, continuously replacing a bifurcat\underline{\smash{\it ing}} regime by a nearby bifurcat\underline{\smash{\it ed}} regime, rendering the whole process, at least in principle, accessible to weakly nonlinear perturbation theory~\cite{mannevref1}.
Open flows with inflectional velocity profiles are prone to an inertial instability (Kelvin--Helmholtz) and, as observed in mixing layers, jets, or wakes, similarly follow a globally supercritical scenario~\cite{mannevref2}.
%\end{minipage}
%\hfill
%\begin{minipage}{0.48\textwidth}
\begin{figure}
%\begin{center}
\centerline{\includegraphics[width=.65\textwidth,clip]{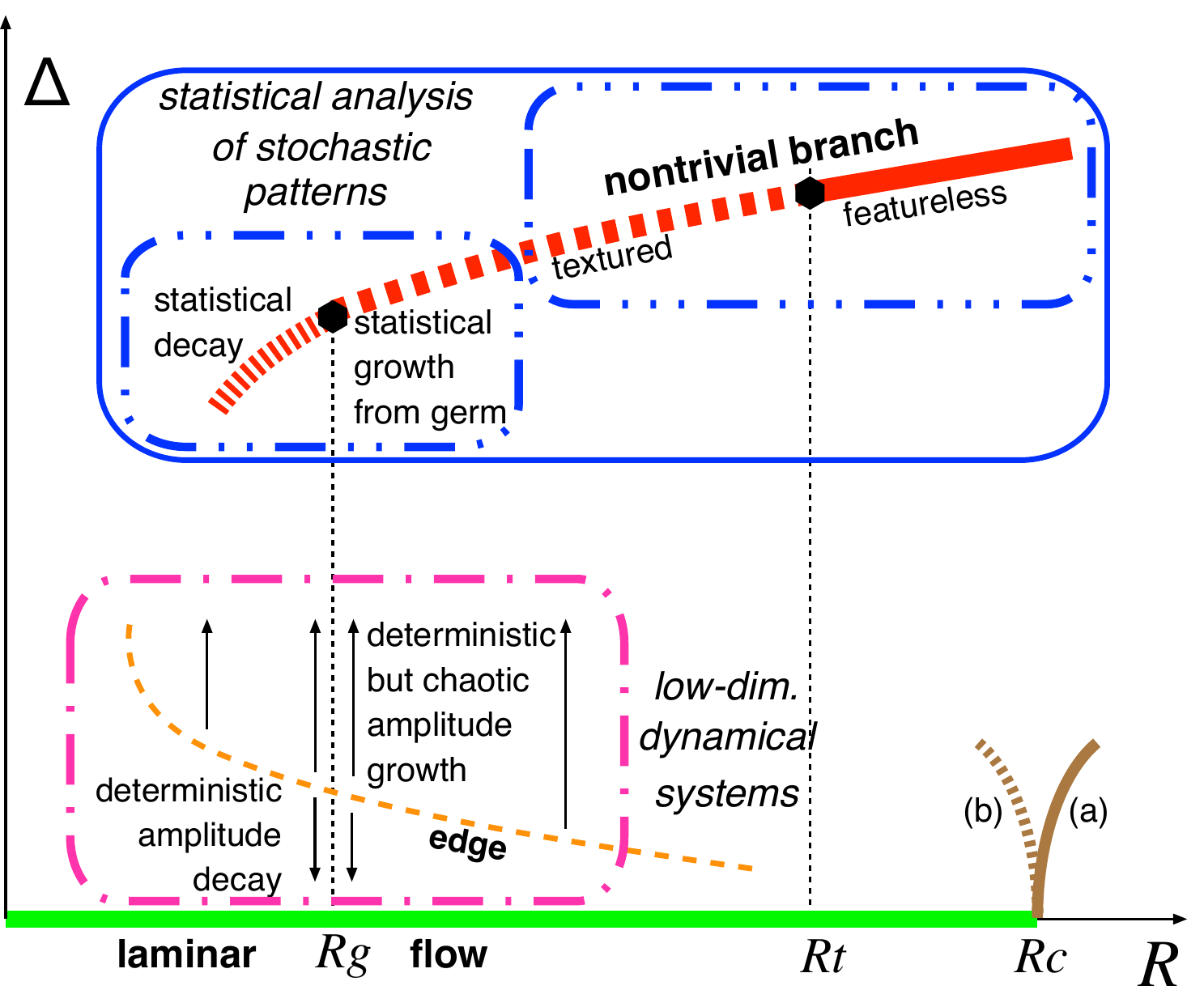}}
%\end{center}

\caption{\label{fig1} Generic bifurcation diagram for transitional wall-bounded flows.}
\end{figure}

This is in strong contrast with the case of wall-bounded shear flows depicted in Fig.~\ref{fig1}.
With velocity profiles deprived from inflection points, unidirectional flows along walls, in tubes or plane channels, boundary layers, etc. remain inertially stable and may experience instability against Tollmien--Schichting waves, but only beyond some high linear threshold $R_{\rm c}$, where $R$ is a suitably defined Reynolds number~\cite{mannevref2}.
$R_{\rm c}$ can even be pushed to infinity, e.g. for flows in ducts of circular or square sections, or for plane Couette flow (PCF).
Whether this instability is supercritical (a) or subcritical (b) [case of plane channel (Poiseuille) flow] is irrelevant since, in practice, much below $R_{\rm c}$, nontrivial states may exist in addition to the trivial laminar state owing to nonlinearities inherent in the advection term of Navier--Stokes equation (NSE).
The flow can therefore ``jump'' to the nontrivial branch populated with chaotic solutions as a result of the amplification of finite amplitude perturbations, hence a direct, discontinuous, {\it globally subcritical\/} transition to turbulence, and the definition of a global stability threshold $R_{\rm g}\ll R_{\rm c}$ below which laminar flow is recovered from any initial condition after a possibly long transient. A general review can be found in \cite{mannevref3}.

Relying on the identification of essential features of the processes sustaining nontrivial states, a cyclical sequence of streamwise vortices inducing streaks by lift-up and subsequent breakdown -- Waleffe's SSP~\cite{mannevref4}, a large body of research on this transition has been focussing on the elucidation of the structure of the phase space in the framework of low-dimensional dynamical systems.
The concept of minimal flow unit (MFU) introduced in~\cite{mannevref5}  --minimal periodized domain in which nontrivial states can persist-- was instrumental in such studies, ending in a picture where unstable coherent structures and manifolds attached to them are key elements~\cite{mannevref6}.
Interesting outcomes of this approach are the interpretation of turbulence's lifetime distributions in terms of chaotic transients and the discovery of {\it edge states\/} sitting on the boundary of the attraction basin of the laminar solution, from the neighborhood of which trajectories can be found to land on the turbulent attractor or else are visited during decay to laminar flow.

By nature unable to deal with spatial modulations in physical space the MFU approach had to be extended to treat the passage from attractor coexistence \underline{\smash{\it in phase space}} to what is actually observed: the coexistence \underline{\smash{\it in physical space}} of domains filled with one of the possible states, either trivial or nontrivial, and interfaces between them, which --by the way-- is generic in extended systems experiencing a subcritical instability.
Accordingly, localized edge states were found, e.g. \cite{mannevref7}, that can serve as germ for the expansion of turbulence above $R_{\rm g}$ or show up at the latest stages of decay.
This directly leads to the interpretation of the transition in terms of spatiotemporal intermittency promoted by Pomeau~\cite{mannevref8} who conjectured that this regime should decay following a scenario interpreted within the phase-transition framework as a transition in the universality class of a stochastic contamination process called {\it directed percolation\/} (DP).
This non-equilibrium process describe the invasion of a uniformly quiescent {\it absorbing\/} regime, here obviously laminar flow, by an {\it active\/} state, here chaotic or turbulent, as the contamination probability increases.
In this framework, the global stability threshold $R_{\rm g}$ corresponds to the DP threshold $R_{\rm c}^{\rm DP}$ and scaling properties are expected in its vicinity.
This statistical-physics viewpoint has received some support from experiments in a quasi-one-dimensional cylindrical Couette flow (CCF) configuration~\cite{mannevref9} and by numerical simulations of a model of shear flow without walls in a quasi-two-dimensional geometry \cite{mannevref10}, among a few other cases.

Channel flow is one such additional case for which agreement has been found with 2D-DP universality in a decay experiment from uniform turbulence  \cite{mannevref11}.
This result is however contradicted by the observation of oblique localized turbulent bands (LTBs) in numerical \cite{mannevref12} or laboratory experiments \cite{mannevref13} below $R_{\rm c}^{\rm DP}$ reported in~\cite{mannevref11}.
Our numerical study, to be developed below~\cite{mannevref14} reveals how these two contradictory results can be reconciled by identifying a transition that preempts the DP scenario just above its supposed critical point.

As $R$ further increases, the spatiotemporally intermittent laminar-turbulent distribution progressively disappears leaving a state of essentially uniform turbulence called {\it featureless\/}~\cite{mannevref16}.
In PCF~\cite{mannevref15} and CCF at large circumferential aspect ratio~\cite{mannevref16,mannevref15}, as well as in channel flow~\cite{mannevref17}, and a few other cases, the laminar-turbulent alternation displays a well-organized oblique turbulent band (or spiral) pattern.
A threshold $R_{\rm t}$ can sometimes be defined for the establishment of the featureless regime, e.g. in PCF \cite{mannevref15,mannevref18}, as also suggested by our preliminary results for channel flow (see below).

Summarizing this long introduction, wall-bounded flows generically experience a direct and wild transition transition to turbulence with different regimes as illustrated in Fig.~\ref{fig1}:
({\it i\/}) nontrivial solution branch disconnected from laminar flow, to be studied using statistical-physics tools in use for critical phenomena, with decay after long chaotic transients below $R_{\rm g}$; 
({\it ii\/}) above  $R_{\rm g}$, regular pattern observable up to an upper threshold $R_{\rm t}$ and featureless turbulent flow above  $R_{\rm t}$;
({\it iii\/}) at the moderate $R$ of interest, viscous effects strong enough to ensure the persistence of coherent structures, making tools from the theory of deterministic, dissipative, low-dimensional dynamical systems still relevant.

Remark also that laminar flow is still a possible {\it linearly stable\/} solution to the NSE all along the transitional range $[R_{\rm g},R_{\rm t}]$ and {\it finite amplitude\/} germs are required to leave the trivial branch, which ordinarily is sufficient to establish the {\it sub-critical\/} character of the transition.
This character is however often expressed trough an expected {\it discontinuous\/} behavior of observables at the corresponding threshold, here $R_{\rm g}$.
This qualitative expectation may be misleading since, quantitatively, the ``distance" between the trivial and nontrivial branch may tend to zero in terms of turbulent fraction, as observed in several cases~\cite{mannevref9,mannevref10,mannevref11}.

\section{Plane channel flow\label{mannevsect2}}

We now turn to our own results for channel flow obtained by direct numerical simulations of NSE using a program developed by one of us.
Details can be found in~\cite{mannevref14}.
The flow is driven by a constant body force $f$.
The centerline velocity $U$ of the laminar flow induced by this force and the half-distance $h$ between the plates are used to turn the equations dimensionless, hence $U=fh^2/2\nu$, with $\nu$ the kinematic viscosity of the fluid, and $R=Uh/\nu=fh^3/2\nu^2$.
A domain $(L_x \times 2 \times L_z)$ is considered with periodic boundary conditions in the stream-wise and span-wise directions ($x,z$) and the usual no-slip condition in the wall-normal direction $y$.
A spectral Fourier--Chebyshev--Fourier scheme is developed in a velocity-vorticity formulation involving the wall-normal velocity $u_y$ and vorticity $\omega_y=\partial_z u_x-\partial_x u_z$~\cite{mannevref19}.
Most simulations have been performed for $L_x=500$ and $L_z=250$ and complementary studies not reported here in four times larger or smaller domains.
The working resolution is with $767$ Fourier wavenumbers in each in-plane direction and $32$ Chebyshev polynomials along $y$.
Aliasing is fully removed by evaluating nonlinear terms using a total of $(2304\times64\times2304)$ mode amplitudes.
The simulations appear to be well-resolved in the range of $R$ of interest to the transition.

\begin{figure}
\includegraphics[angle=90,width=0.19\textwidth]{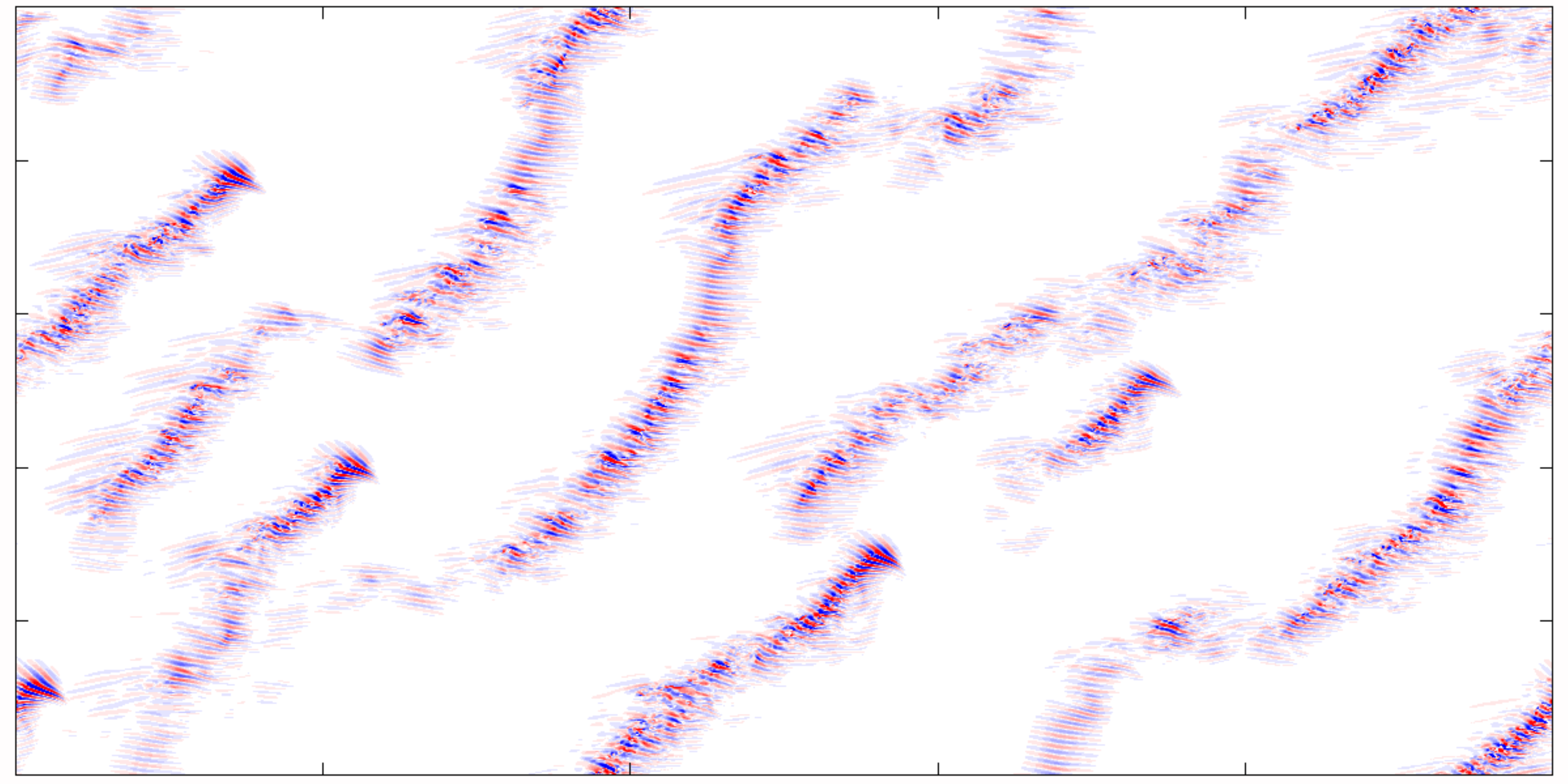}\hfill
\includegraphics[angle=90,width=0.19\textwidth]{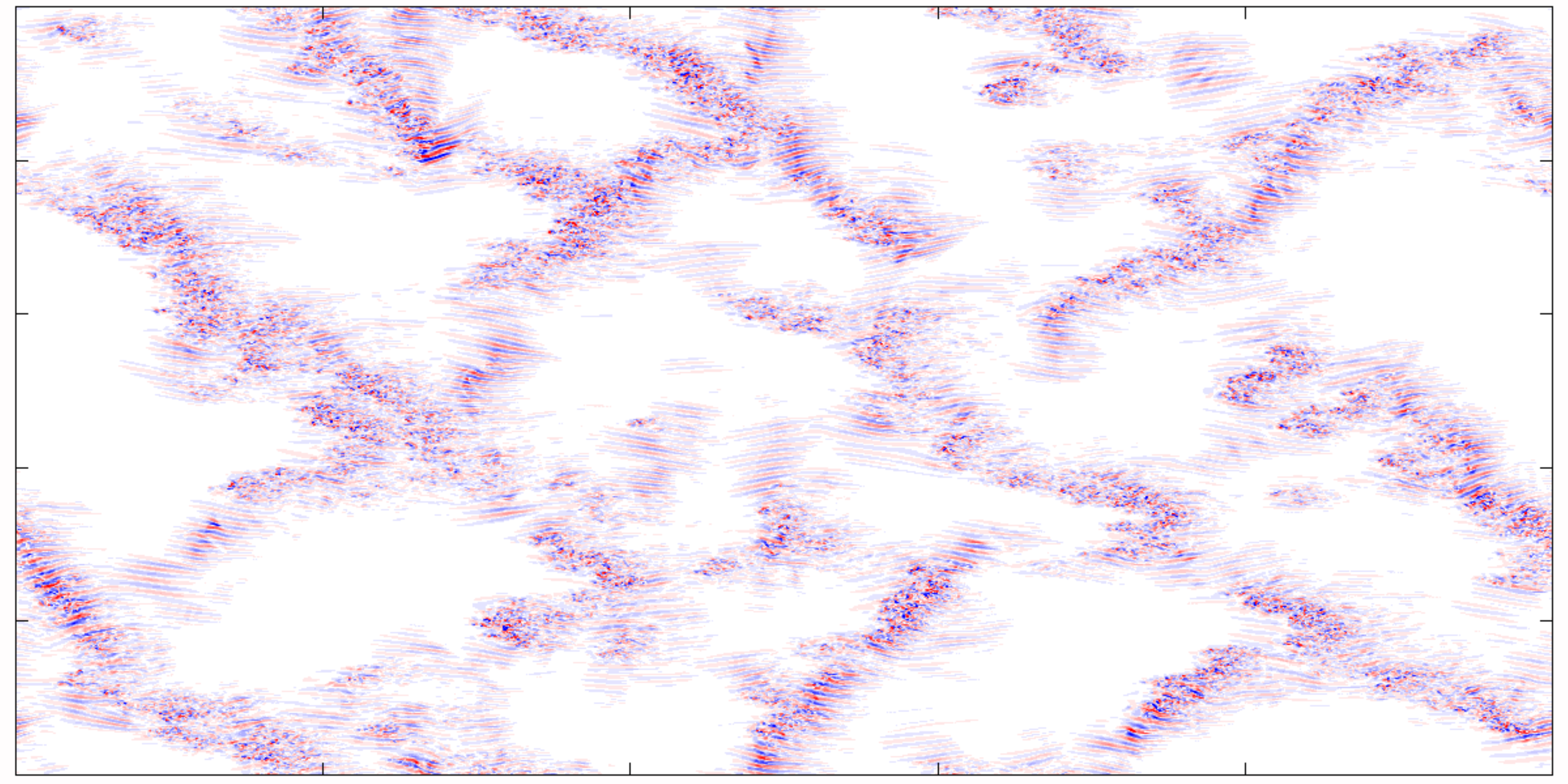}\hfill
\includegraphics[angle=90,width=0.19\textwidth]{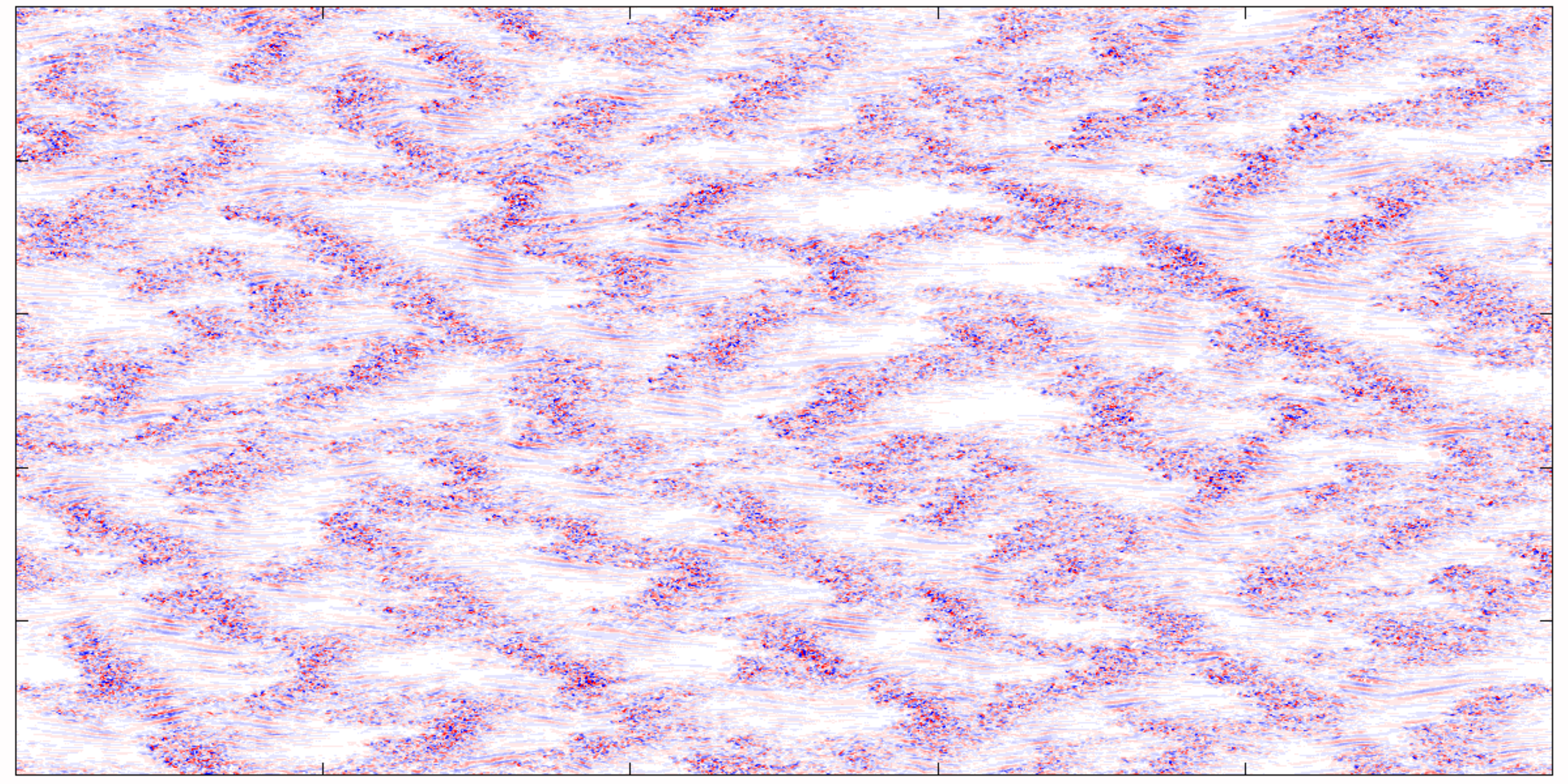}\hfill
\includegraphics[angle=90,width=0.19\textwidth]{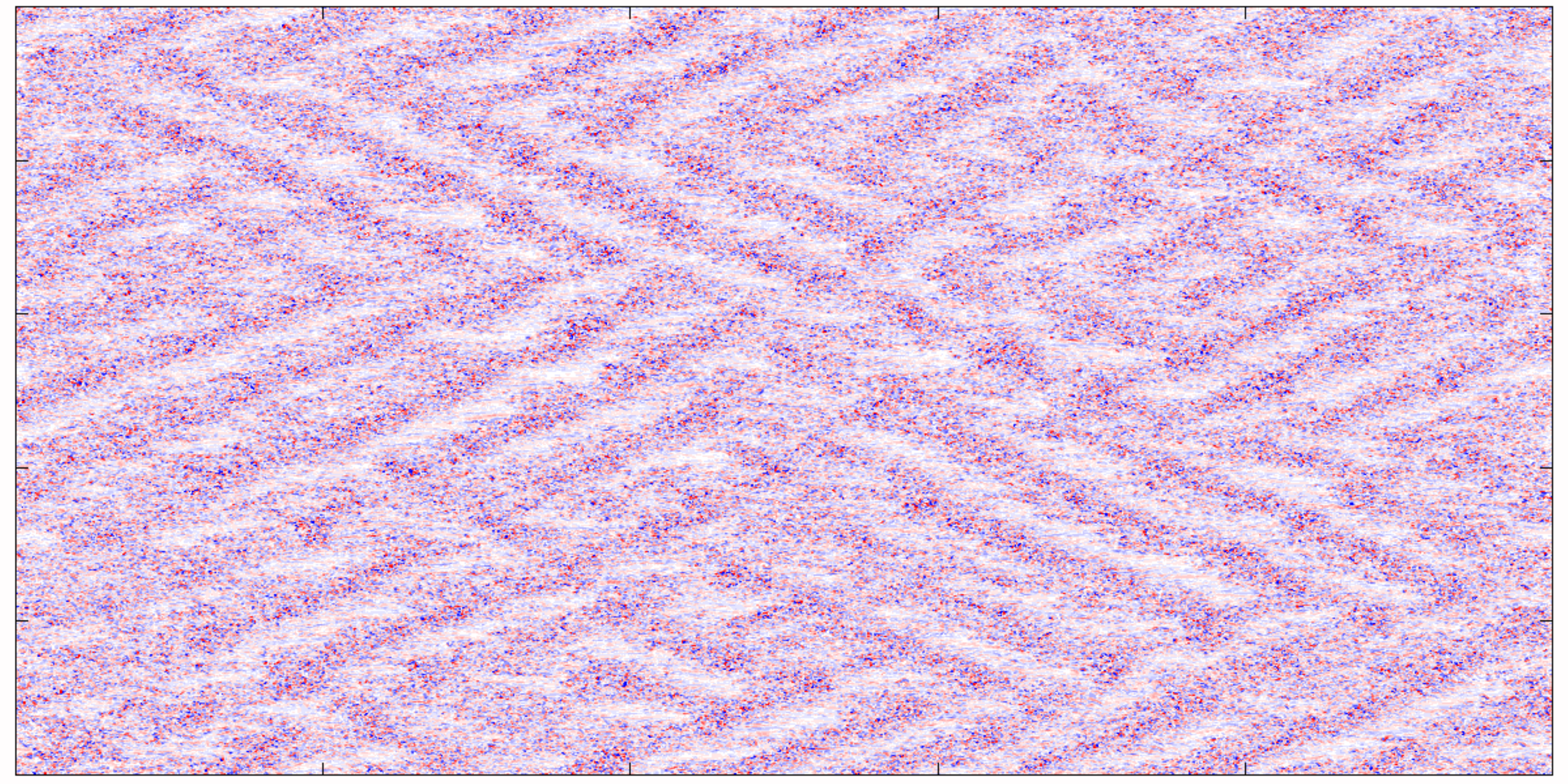}\hfill
\includegraphics[angle=90,width=0.19\textwidth]{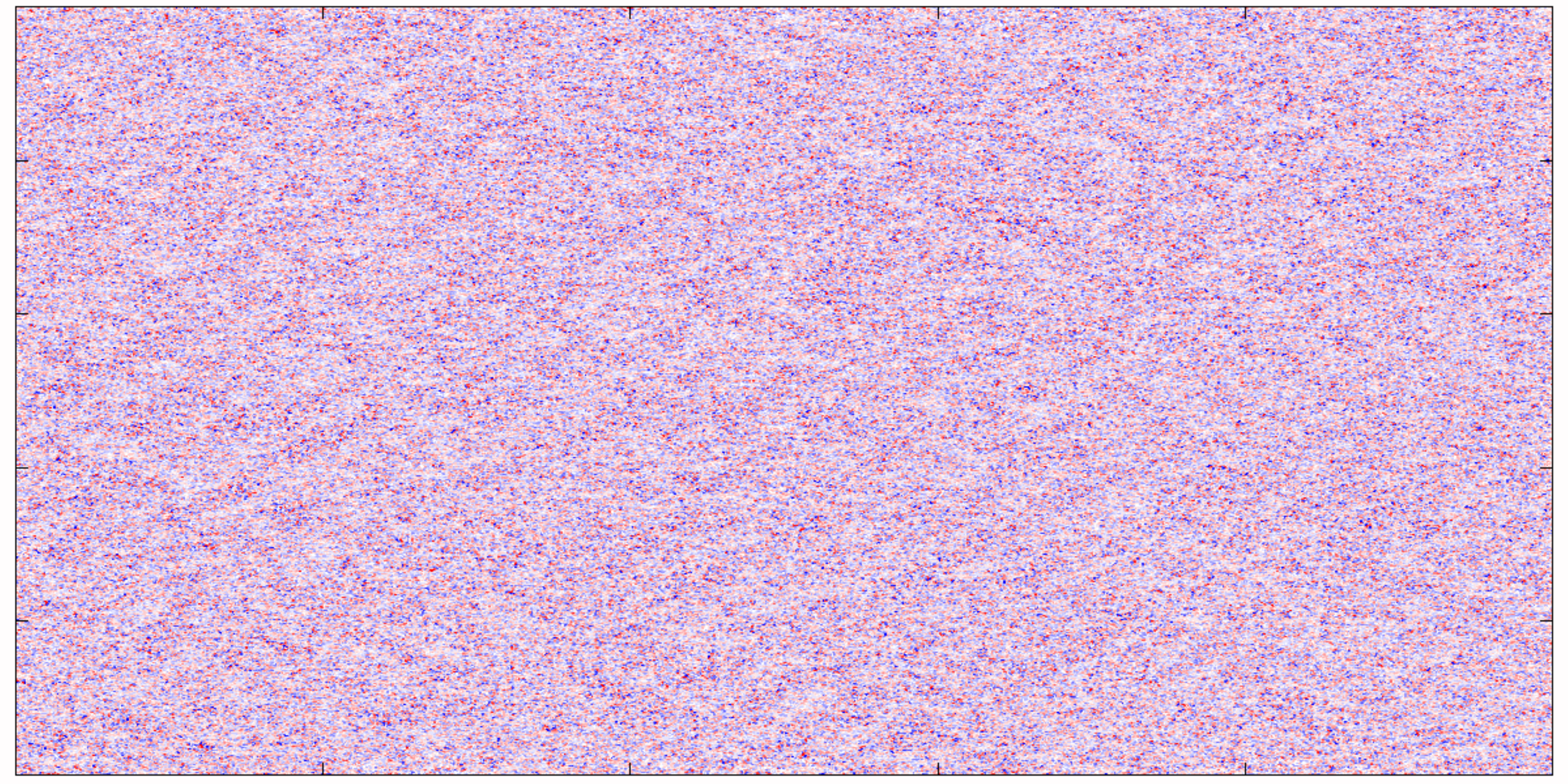}

\footnotesize $\null\hspace*{0.7em}R=850,\tilde R =789 \hspace{1.7em} R=1200,\tilde R=1012 \hspace{1.1em} \hfill R=1800,\tilde R=1237 \hspace{1.em} \hfill R=3000,\tilde R=1604\hspace{1em} \hfill R=4000,\tilde R=1843\hspace{0.2em}$

\caption{\label{fig2} Statistically steady regimes of channel flow as depicted from the wall-normal velocity component measured in the central plane in color levels (white is laminar). Stream-wise direction is vertical upwards.}
\end{figure}

\paragraph{Overview of the transitional range in physical space.}
At given $R$ below $R_{\rm c} =5772$~\cite{mannevref20}, besides the laminar solution, NSE possess solutions triggered by finite perturbations and belonging to the nontrivial branch mentioned in \S\ref{mannevsect1}.
Such solutions are associated to a mean stream-wise velocity $\langle U_{\rm m}\rangle$ smaller than the corresponding laminar value $U_{\rm m}^{\rm lam}=\frac23$ (in units of $U$).
Due to turbulence, speed $\langle U_{\rm m}\rangle$, often called {\it bulk velocity\/}, is an empirical quantity that fluctuates and has to be averaged in time.
Accordingly, we define $\tilde R = \frac32 \langle U_{\rm m}\rangle R$ to facilitate the comparison with other works that use the bulk velocity in the definition of the Reynolds number ($\tilde R\equiv R$ for laminar flow).
The different flow regimes we have observed are depicted in Fig.~\ref{fig2}.
A preliminary simulation at $R=2000$, starting from a finite amplitude initial random condition, has provided us with a pattern that has next been evolved upon decreasing or increasing $R$ regularly (values up to $R=6000\gtrsim R_{\rm c}$ have been considered).
Solutions presented have reached a statistically steady state. 

LTBs described in~\cite{mannevref12,mannevref13} are here shown for $R=850$, in the one-sided propagation regime where all the bands go in the same direction.
They are seen to recede and disappear for $R\lesssim 700$.
At $R=1200$ they propagate in both directions, forming a strongly intermittent loose discontinuous laminar-turbulent network, with turbulent arms broken by laminar gaps.
At $R=1800$, laminar gaps have disappeared and the network is now continuous but still loose and intermittent.
When $R$ is further increased this crisscrossed pattern gets tight and more regular with conspicuous domains of one or the other orientation separated by grain boundaries, here at $R=3000$.
This pattern next fades away, being hardly visible at $R=4000$.
A close parallel can be drawn with the case of CCF (and PCF) examined by the Saclay group~\cite{mannevref15}.

\paragraph{The lower transitional range.}
Below $R\sim 3000$, the laminar-turbulent alternation is sufficiently marked for the definition of a turbulent fraction $F_{\rm t}$ to make sense.
This turbulent fraction was determined using a two-level moment-preserving thresholding method~\cite{mannevref20} that automatically places the cut-off between laminar and turbulent local states so as to reduce the distribution of the gray-levels to an optimal black-and-white distribution. 
\begin{figure}
\begin{center}
\includegraphics[width=0.49\textwidth]{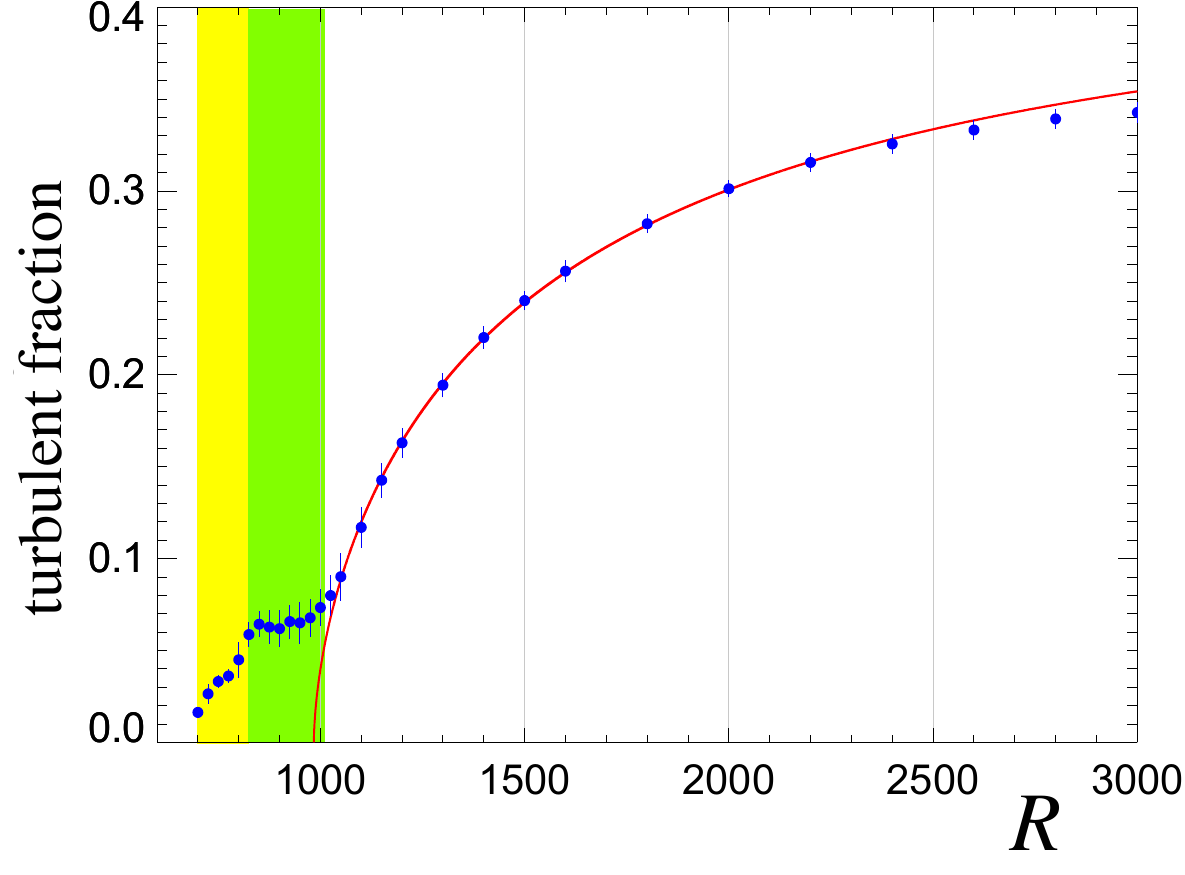}\hfill
\includegraphics[width=0.49\textwidth]{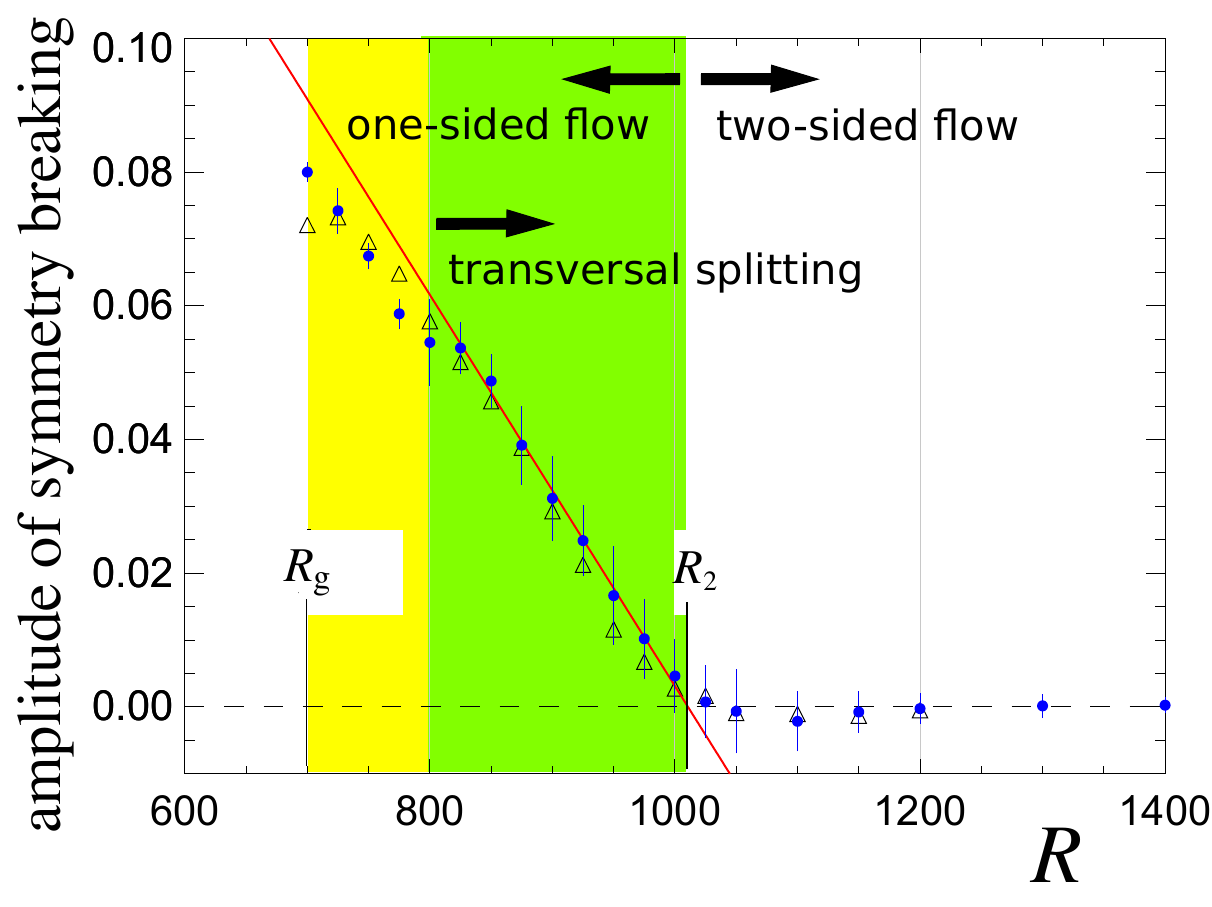}
\end{center}
\vspace*{-2ex}

\caption{\label{fig3}Left: Turbulent fraction as a function of $R$. Right: Symmetry breaking near the one-sided/two-sided bifurcation point at $R_2$; triangles are for simulations in a four-fold wider domain.}
\end{figure}

Figure~\ref{fig3} (left) displays the variation of $F_{\rm t}$ as a function of $R$.
The line corresponds to a fit of the data against the function$\!$%
\footnote{The control parameter is $1/R$ rather than $R$, yielding an excellent fit from $R=1050$ up to $2400$.} 
 $F_{\rm t}= A (1-R_{\rm c}^{\rm DP}/R)^\beta$ with $A\simeq0.45$,  $R_{\rm c}^{\rm DP}\simeq984$, and $\beta \simeq0.58\approx \beta^{\rm DP}$ in the two-dimensional case.
 The behavior reported by Sano \& Tamai~\cite{mannevref11} is therefore recovered but the DP critical point cannot be closely approached since for $F_{\rm t} \lesssim 0.1$ a crossover is observed involving a symmetry-breaking bifurcation from two-sided to one-sided propagation that precisely corresponds to the LTB regime observed in~\cite{mannevref12,mannevref13} at comparable Reynolds numbers.

At decreasing $R$, the symmetry-breaking transition is due to the rapidly decreasing probability of lateral branching compared to splitting along the LTBs.
 The two orientations appear with essentially equal weights and the rate of lateral branching is comparable to that of parallel splitting in the symmetrical regime ($R=1200$ in Fig.~\ref{fig2}).
But this rate rapidly decreases in the green zones appearing in Fig.~\ref{fig3} and becomes negligible for $R\le800$ in the yellow zones.
An observable measuring the lack of symmetry has been designed as plotted in Fig.~\ref{fig3} (right). 
A simple phenomenological model has been developed for this symmetry restoration~\cite{mannevref14}, supporting a linear variation of the asymmetry amplitude that helped us extrapolate it to zero and define the corresponding threshold at $R_2\simeq1010$.

\paragraph{The upper transitional range.}
For $R\gtrsim 2600$, the spatiotemporally intermittent laminar-turbulent alternation is less marked rendering the analysis in terms of turbulence fraction less appropriate while the pattern becomes both more steady and more regular, suggesting the recourse to Fourier analysis.
The transverse perturbation energy field $E_{2D}(x,z;t) = \int_{-1}^{+1} {\rm d} y \frac12 [u_y^2(x,y,z;t)+u_z^2(x,y,z;t)]$ is considered here and series of Fourier spectra are recorded at statistically steady state.
Means and standard deviations of time series of the intensity of each mode with $k_{x,z}=2m_{x,z}\pi/L_{x,z}$, i.e. $S_{k_x,k_z;n}=| \hat E_{\rm 2D}(k_x,k_z;t_n)|^2$, are computed for  50 snapshots separated by $\Delta t=1000$, hence essentially independent.

\begin{figure}
\begin{center}
\includegraphics[width=0.32\textwidth]{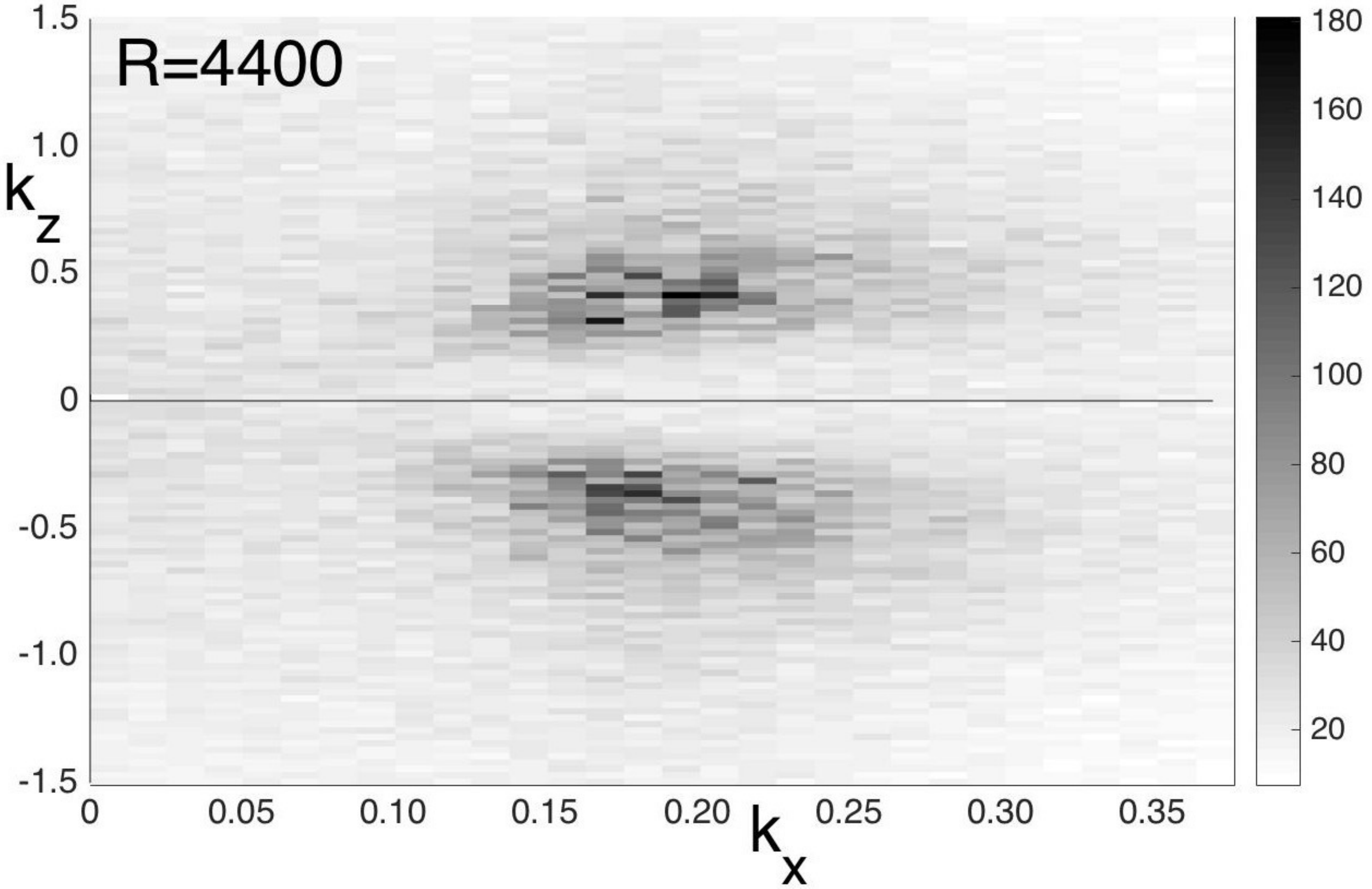}\hfill
\includegraphics[width=0.32\textwidth]{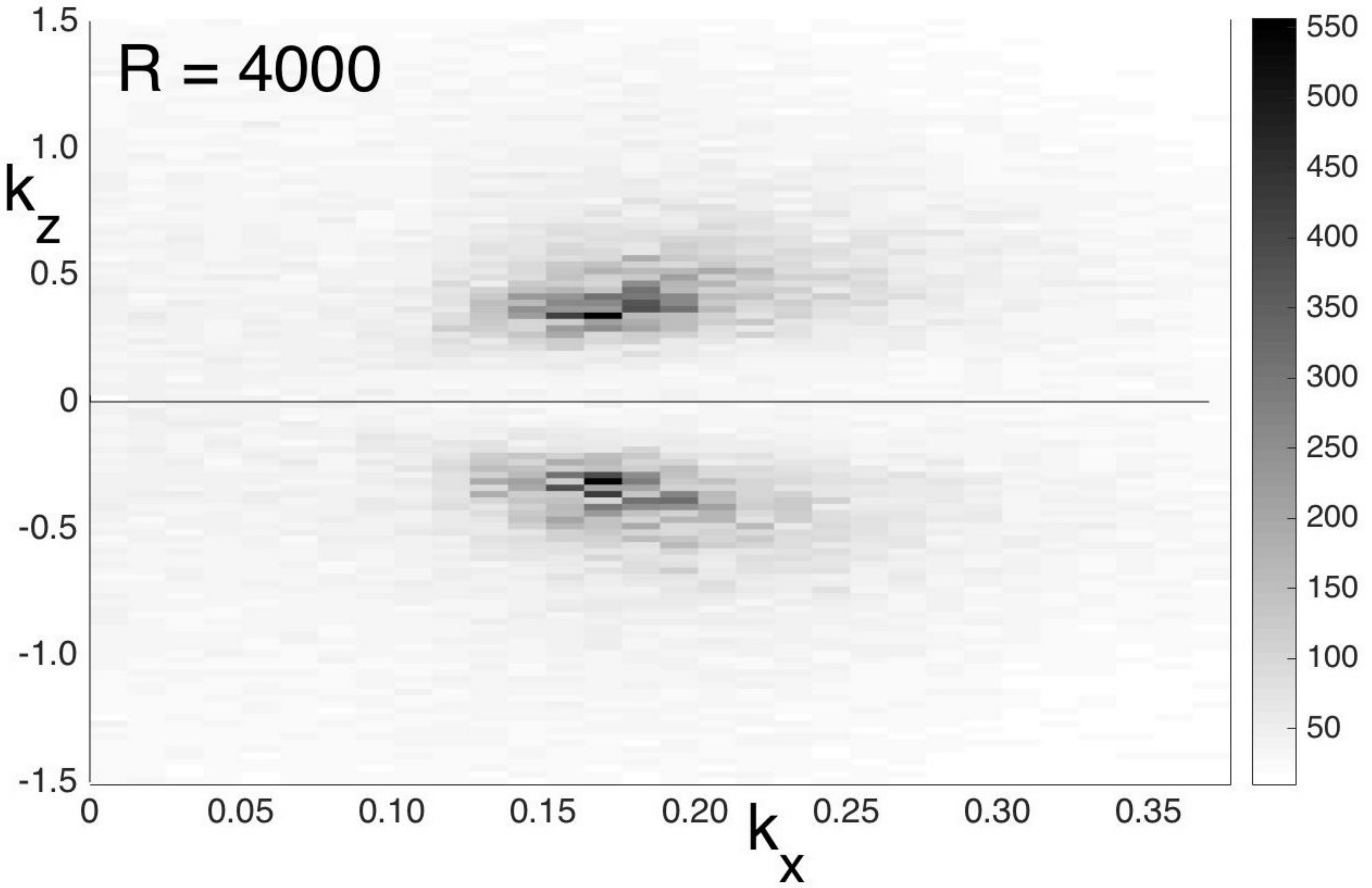}\hfill
\includegraphics[width=0.32\textwidth]{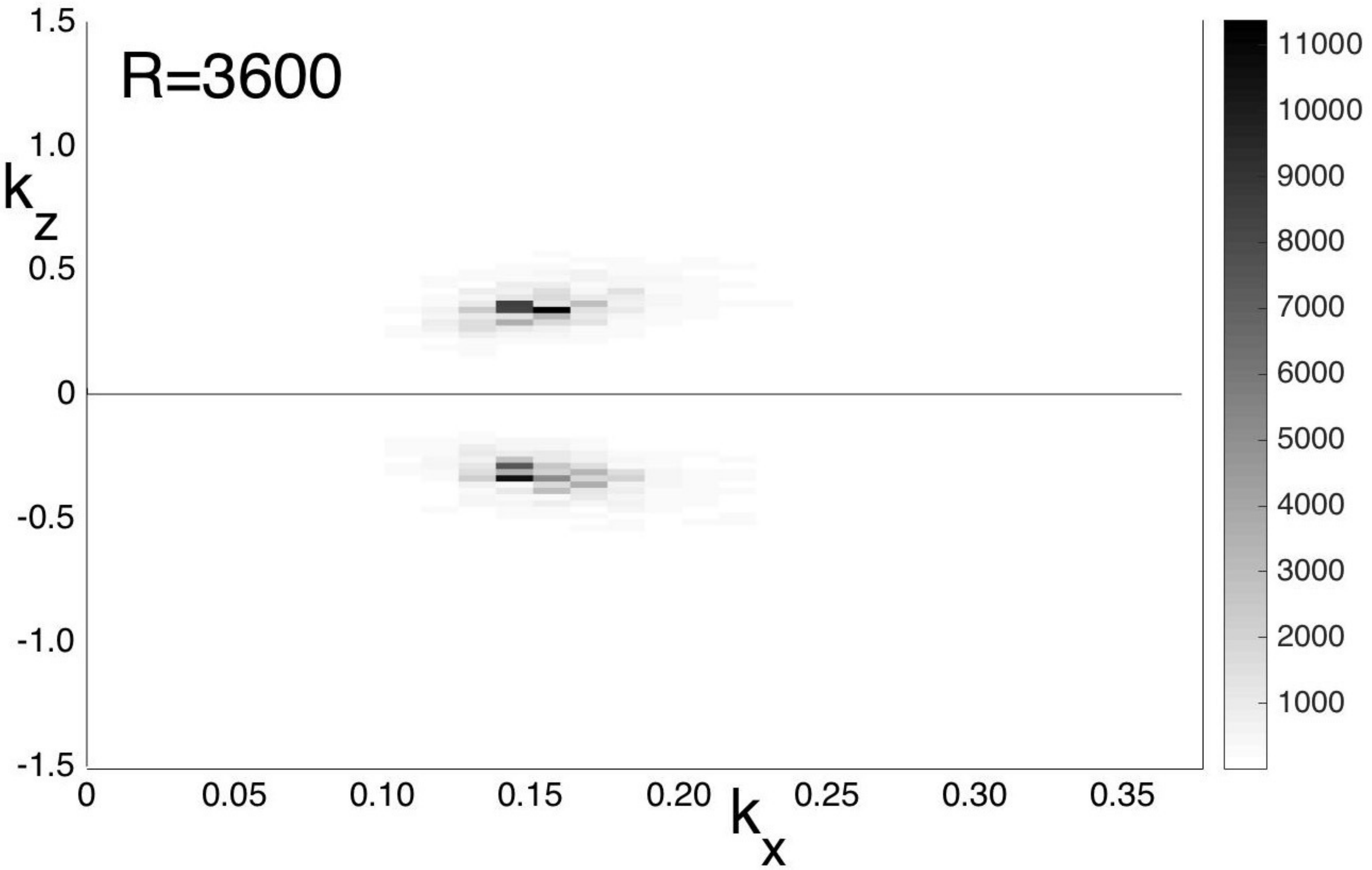}\\[2ex]
\includegraphics[width=0.48\textwidth,clip]{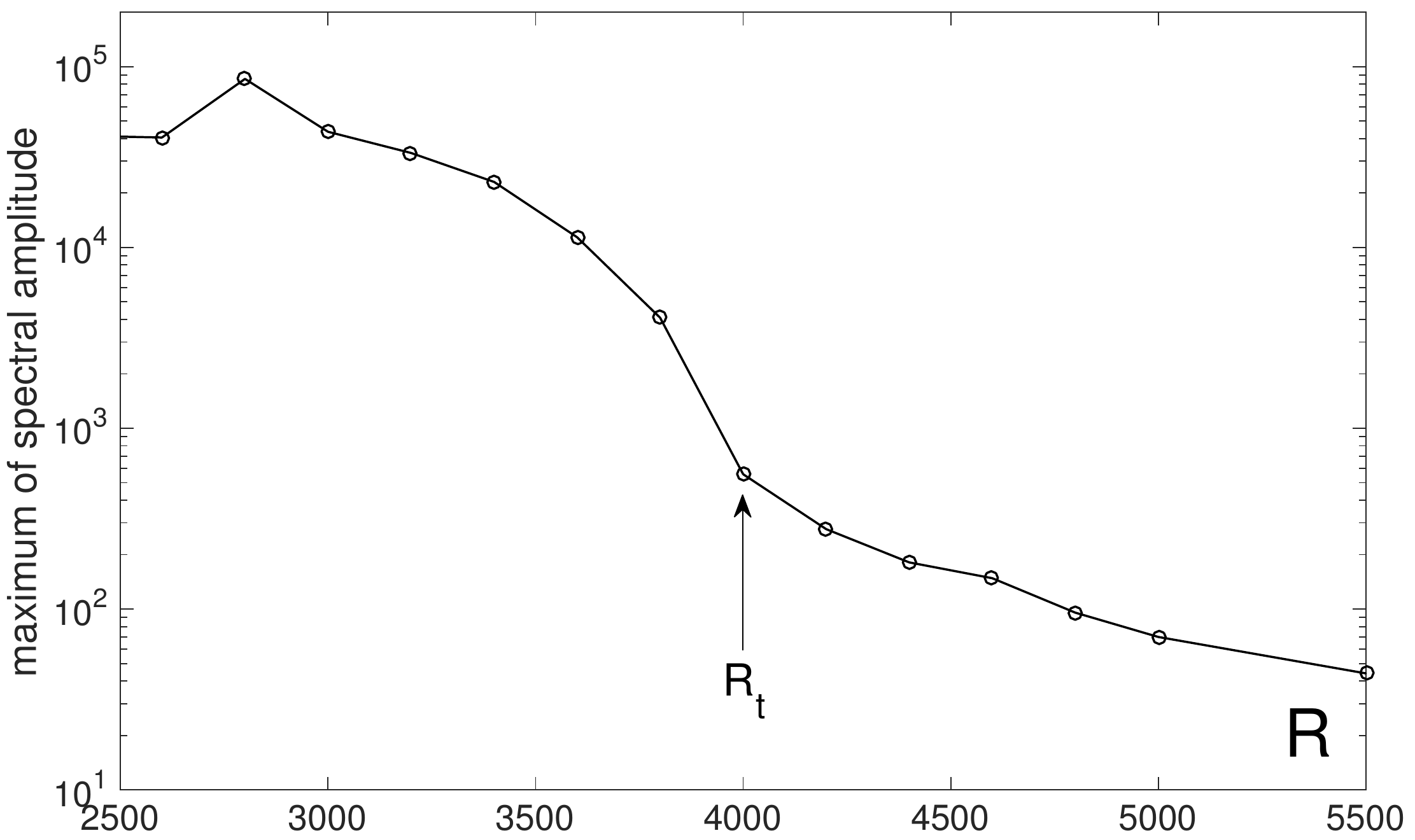}\hfill
\includegraphics[width=0.48\textwidth,clip]{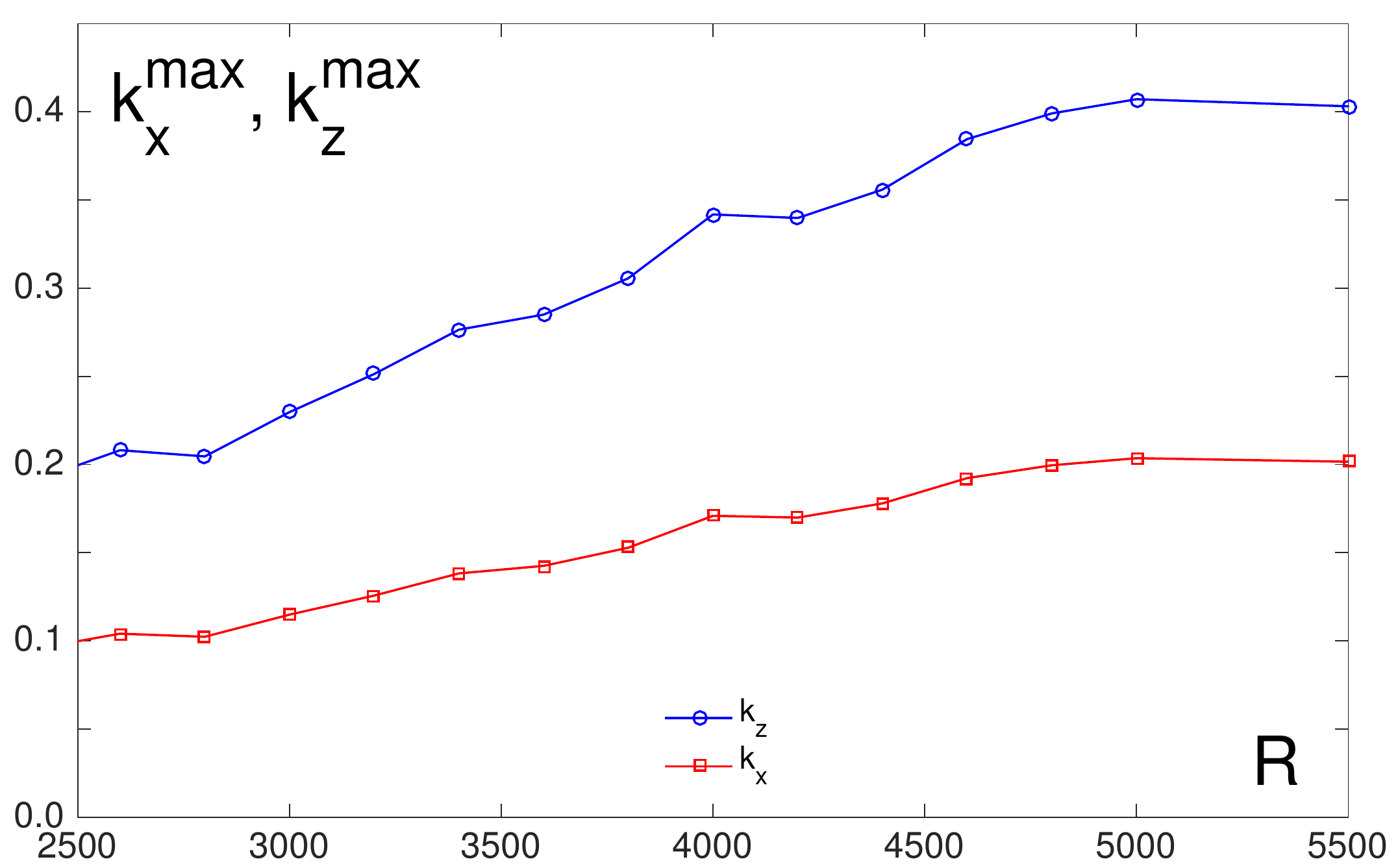}
\end{center}

\caption{\label{fig4}Top: Mean spectra $\overline S_{k_x,k_z}$ of $E_{2D}$ for $R=4400$, $4000$, and $3600$, from left to right.
Bottom. Maximum of spectral intensity $\overline S_{k_x,k_z}^{\rm max}$ (left) and location $k_x^{\rm max},k_z^{\rm max}$ of this maximum in the ($k_x,|k_z|$)-plane (right), as functions of $R$.}
\end{figure}

Figure~\ref{fig4} (top) display the means of these spectra $\overline S_{k_x,k_z}$.
At large $R$ inside the supposedly {\it featureless\/} regime, wide bumps are found, corresponding to highly fluctuating, low amplitude, modulations, less and less pronounced as $R$ increases ($\overline S^{\rm max}(4400)\simeq 180$).
In contrast, when the pattern is well established the spectra are dominated by sharp peaks ($\overline S^{\rm max}(3600)\simeq 11200$).
The variation of the maximum $\overline S^{\rm max}$ as a function of $R$ is displayed  using lin-log scale in Fig.~\ref{fig4} (bottom, left).
It shows that the inflection at $R\sim 4000$ ($\overline S^{\rm max}(4000)\simeq550$) corresponds to a rapid increase, interpreted as a sudden condensation of the weak turbulence-intensity modulations into a genuine pattern.
Figure~\ref{fig4} (bottom, left) displays the position $(k_x^{\rm max},k_z^{\rm max})$ of these maxima in the wave-vector plane as a function of $R$.
It is seen that for $R\in [2500,4000]$, the two-components of the wave vectors increase roughly linearly with $R$, i.e. a regular wavelength decrease from $(\ell_x,\ell_z)\simeq (62,32)$ at $R=2500$ to $\simeq(38,19)$ at $R=4000$.
They however maintain a nearly constant ratio $|k_z|/k_x \sim 2$, hence turbulent bands essentially making an angle $\theta=\arctan(1/2) \approx26^\circ$ with the streamwise direction all along this range.
For $R\ge4000$, the wavelength of the turbulence-intensity fluctuations continues to decrease slightly as they become weaker and weaker with increasing $R$. 

\section{Discussion/Conclusion\label{mannevillesect3}}
The transition to turbulence in wall-bounded flows, briefly reviewed in \S\ref{mannevsect1}, remains a good testing ground for the theory of non-equilibrium phase transitions.
In \S\ref{mannevsect2}, we illustrated the case of  plane channel flow driven by a constant bulk force using our own simulations.
Results presented in the recent literature, among others~\cite{mannevref11,mannevref12,mannevref13,mannevref17}, were recovered.
In particular, the relevance of the 2D-DP scenario for turbulence decay \cite{mannevref11} was confirmed but reconciled with other results showing the presence of localized turbulent states below the DP-threshold~\cite{mannevref12,mannevref13}.
Within our protocol, a symmetry-breaking bifurcation due to the rapid decrease of the rate of lateral branching in the loose spatiotemporally intermittent network regime forces the system to a new regime preempting the critical behavior inherent in the DP framework~\cite{mannevref14}.
Ongoing work is devoted to improve the statistical study of the flow in the upper transitional range.
The physical mechanisms underlying the branching processes at increasing $R$ near $R_{\rm g}$ and those responsible for the patterning observed at decreasing $R$ at the still putative threshold $R_{\rm t}$ remain to be elucidated. 

\paragraph{Acknowledgments.}
This work is supported by JSPS KAKENHI Grant Number JP17K14588. Simulations are performed using NIFS's FUJITSU FX100 ``Plasma Simulator'' (contract NIFS16KNSS083).
We would like to thank Y. Duguet (LIMSI) for interesting discussions about the problem.

\end{document}